\documentclass[10pt,a4paper]{article}

 \usepackage[dvips]{graphics}
 \usepackage{graphicx}
 \usepackage{afterpage}
 \usepackage{enumerate}

\begin{document}

 \title{\bf Temperature stratification\\ of the atmosphere of Arcturus
}

 \author{\bf V.A. Sheminova}
 \date{}

 \maketitle
 \thanks{}
\begin{center}
{Main Astronomical Observatory, National Academy of Sciences of
Ukraine
\\ Zabolotnoho 27, 03689 Kyiv, Ukraine\\ E-mail: shem@mao.kiev.ua}
\end{center}

 \begin{abstract}
A brief overview of the results of the investigations  of the red giant star
Arcturus is given. One-dimensional LTE modeling of the atmospheres of
Arcturus and the Sun as a star is carried out on the basis of synthesis of
the extended wings of the H and K Ca II lines.  It is found that the local
continuum in this spectral region is underestimated by an average of 12\% in
the atlases of Arcturus. The average deficit in UV absorption amounts to
43\% for Arcturus  whereas it is  9\% for the Sun. For Arcturus the
correction factor to the continuum opacity  at the wavelengths of 390.0,
392.5, 395.0, 398.0, and 400.0 nm equals 2.20, 1.90, 1.70, 1.55, and 1.45.
The model atmosphere of Arcturus obtained from the best-fit of the wings of
the H and K Ca~II lines corresponds to the model atmosphere with the 
fundamental parameters $T_{\rm eff} = 4286$~K,  $\log g = 1.66$, and
[Fe/H]=$-$0.52 derived by Ramirez and Allende Prieto (2011). The 
temperature stratification of Arcturus' atmosphere is presented in tabular form. 
To obtain more accurate temperature stratification in the future, 
we need a high spectral resolution spectrum calibrated to absolute 
fluxes with high accuracy.

\end{abstract}

\section{Introduction}
     \label{S-Introduction}

Arcturus ($\alpha$ Boo, HR5340, HD124897, HIP69673) is a red giant of a K1.5
IIIp spectral type with an age $7_{-1.2}^{+1.5}$ of billion years (Ramirez
and Allende Prieto \cite{56}).
It is 110 times visually brighter than the Sun, and its total luminosity
exceeds the solar luminosity by a factor of 180. Arcturus is the second
brightest star of the northern sky ($V=-0.051\pm 0.013^m$, $M_v=-0.313\pm
0.016^m$) and is located at a distance of 36.7 ly
or $11.26_{-0.07}^{+0.05}$~pc (Perryman et al. \cite{53}) from the earth.
Its parallax is equal
to $0^{\prime\prime}.08883\pm 0^{\prime\prime}.00053$ (van Leeuwen \cite{75}).
Arcturus is
classified as a variable star, and the ``Suspected Variable Stars'' section
of the General Catalogue of Variable Stars (GCVS,
http://www.sai.msu.su/gcvs/gcvs/) lists two values ($ V_{\rm max}= -0.13^m$
and $\rm V_{\rm min}= -0.03^m$) of its apparent magnitude. Arcturus is a
pulsating star that exhibits complex multiperiodic variations of radial
velocity. It is believed that the short-period (1.8, 2.5, 4.0, 8.3, and 46.0
days) variability is caused by radial pulsations and oscillations
(Hatzes and  Cochran \cite{31}),
whereas the two-year periodicity is caused by modulation due to rotation
(Gray and Brown \cite{24}). A magnetic activity cycle with a period
exceeding 14 years was recently found by  Brown et al. \cite{15}.

A relatively low declination ($+19^\circ$) makes it possible to observe
Arcturus from both hemispheres of the earth. Its proper motion and radial
velocity are equal to $2.3^{\prime\prime}$ per year and $-5$~km/s,
respectively, while its velocity of space motion relative to the Sun equals
120~km/s. Its high spatial velocity and low metallicity ([A/H]~$=-0.5$)
indicate that Arcturus belongs to a group of old stars residing in the thick
disk of the Galaxy in the neighborhood of the Sun (Ramirez
and Allende Prieto \cite{56}). The discussed
star is a member of the Arcturus kinematic group (Eggen \cite{19}),
 which consists
of 53 stars. It is assumed that this group is extragalactic in origin.
According to  Navarro et al. \cite{48}, all characteristics of the Arcturus group are best
matched by the characteristics expected from a group of stars that are left
over from a small galaxy that was destroyed and devoured by the Milky Way.
It is the devoured galaxies that could introduce a significant number of old
stars with low metallicities into the disk of the Milky Way. This exotic
assumption certainly requires further assessment.

The atmospheric parameters of Arcturus were determined multiple times using
different methods. This star attracts constant interest of astrophysicists
due to the fact that it may well be used as a standard for spectroscopic
studies of red giants. A differential analysis of the giants with respect to
Arcturus makes it possible to minimize the systematic errors of estimation
of the chemical composition and fundamental parameters  (Alves-Brito et al.
\cite{6}).

The most complete review of the results of the studies of the atmosphere of
Arcturus was presented by Trimble and  Bell \cite{71}. An insufficiently
accurate value of the star's mass still remains the main problem of these
studies. Since observations of the star's secondary component are lacking
thus far, it is not possible to determine the mentioned mass directly. For a
long time, it was widely believed that Arcturus is a single star, but
relatively recent observations carried out by the space astrometry satellite
HIPPARCOS (Perryman et al.\cite{53}) revealed the binarity of Arcturus (this result was
labeled ``unreliable''). According to Perryman et al. \cite{53},
the secondary component is
  \begin{figure}
    \centering
   \includegraphics[width=13 cm]{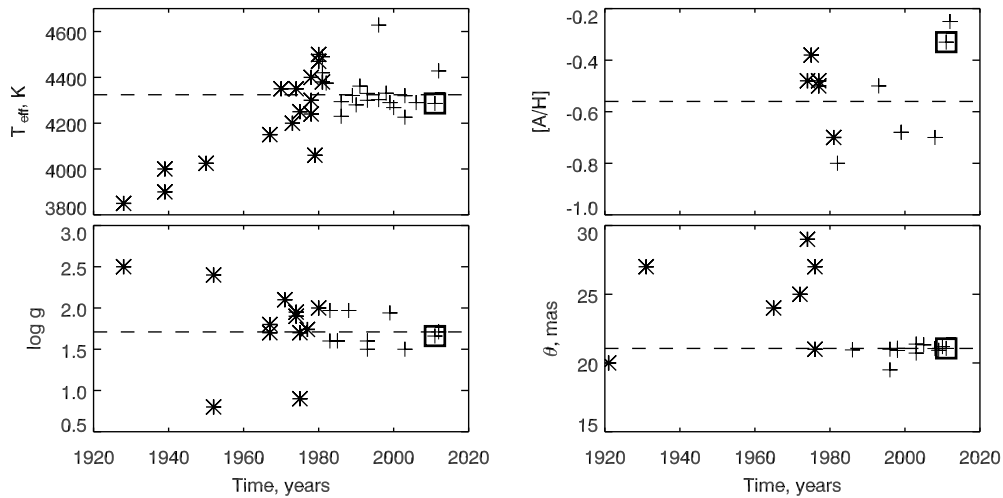}
   \caption
{Different estimates of the effective temperature $T_{\rm eff}$, surface
gravity acceleration $\log g$, metallicity [A/H], and angular diameter Þ of
Arcturus that were presented over the years. Data prior to 1981 (asterisks)
are taken from  the review of Trimble and  Bell \cite{71}. Recent results
Ramirez and  Allende Prieto \cite{56} are marked by squares.
The dashed line denotes average values taken from the PASTEL database. }
  \label {Fig1}
  \end{figure}
weaker (as follows from assessment of the spectrum around $\lambda =
460$~nm) than Arcturus by $3.33\pm0.31^m$ and is located at a distance of a
mere $0^{\prime\prime}.255\pm0^{\prime\prime}.039$ from Arcturus. Consequent
ground observations with a 100-inch Mount Wilson telescope, which were
carried out using an adaptive optics system in the visible range (Turner et
al. \cite{73}), did not confirm the binarity of Arcturus. The interest in
this intriguing problem was renewed recently when researchers were trying to
interpret the results of interferometric observations in the near IR range
(Verhoelst et al.\cite{76}). Systematic differences between the observations
and the calculations were successfully resolved only on the assumption of
binarity, the secondary component of the suspected binary system being
classified as a subgiant of a G spectral type. Ramirez and  Allende Prieto
\cite{56} suggest that the binarity of Arcturus may explain the excess flux
of radiation in the UV range that is predicted by modern models of the
atmosphere of Arcturus. At present, the question of the binarity of Arcturus
remains unsettled. We consider Arcturus to be a single star.

A new database of the effective temperature $T_{\rm eff}$, surface gravity
acceleration $\log g$, and metallicity [A/H] values, which is presented in
the PASTEL (Chemin et al. \cite{17}) catalogue, testifies to the fact that
the number of papers dealing with determination of the atmospheric
parameters of Arcturus continues to grow. The spread of $T_{\rm eff}$ and
$\log g$ values has lowered significantly in the past decades (see Fig.~1).
The values of $T_{\rm eff}$ have stayed within the range of 4226 to 4362~K
since 1982 if we exclude one interferometric measurement by Dyck et al.
\cite{18} that gives a high value of $4628 \pm 133$~K. The values of $\log
g$ stay within the range of 1.50 to 1.94. The spread of the [A/H] values has
seen no substantial reduction in these years, though results of the
determination of metallicity using high-resolution spectra were presented in
more than 20 papers. The metallicity value varies from $-0.25$ to $-0.80$.
The accuracy of the angular diameter $\theta$ measurement has increased
considerably. The average error of its measurement amounts to
$0^{\prime\prime}.0002$.

The results obtained in the studies of the fundamental parameters of
Arcturus were summarized recently by Ramirez and  Allende Prieto \cite{56}.
They also carried out a new spectral analysis of Arcturus  on the basis of
the most reliable data and techniques, and obtained the values of $T_{\rm4
eff} = 4286 \pm 30$~K, $\log g = 1.66 \pm 0.05$, [Fe/H]~$ =- 0.52 \pm 0.04$,
$M = (1.08 \pm 0.06)M_{\odot}$, and $R = (25.4 \pm 0.2)R_{\odot}$, as well
as the elemental abundances based on the equivalent widths of the spectral
lines. These values of the fundamental parameters agree within the limits of
error with the average values ($T_{\rm eff} = 4324 \pm 90$~K, $\log g = 1.71
\pm 0.29$, and [A/H]~$ =-0.56 \pm 0.10$) derived from PASTEL (Chemin et al
\cite{17}).

The distribution of temperature with depth in the atmospheres of stars is
derived either theoretically (on the basis of the constancy of radiation
flux at each depth value) or using the scaling factor that is applied to the
solar data. A semiempirical approach that is based on observations of the
distribution of energy in the star's spectrum or of a multitude of chosen
spectral lines with a high resolution is used less often. The theoretical
approach prevails. The computed theoretical models possess no free
parameters. The overall opacity is calculated with account for all kinds of
absorption sources, including all absorption lines. The full list of
absorption lines, which is needed to accurately calculate the opacity, is
constantly updated (e.g., Grupp et al. \cite{27}). Nowadays, this list
numbers millions of atomic lines, which are stored in the VALD
(http://www.astro.uu.se/~vald/php/vald.php) database (Piskunov et al.
\cite{55}) together with the atomic parameters. A list of molecular lines of
Jorgensen and Lindegren \cite{39} is also constantly updated. After the
opacity calculation, the temperature and pressure stratification, which is
defined by the energy conservation law, is determined. The ATLAS (Kurucz
\cite{44}) and MARCS (Gustafsson et al. \cite{30}) software packages are
most frequently used for computing the theoretical models of stellar
atmospheres. These packages are constantly improved in order to cover an
increasingly broad class of stars. The MARCS package was recently adapted to
the Linux operating system by Sbordone et al. \cite{59} and is now freely
available. The grids of theoretical models with a broad range of values of
the main parameters ($T_{\rm eff}$, $\log g$, [A/H], and the microturbulent
velocity $V_{\rm mic}$) were computed using ATLAS and MARCS. A model of the
atmosphere of a given single star may easily be obtained by interpolating
several models taken from the ATLAS (http://kurucz.harvard.edu/) or MARCS
(http://marcs.astro.uu.se/) grids.

It should be noted that the theoretical modeling of physically
self-consistent models has enjoyed a significant advancement. Traditional
approximations of plane-parallel stratification in the horizontally
homogeneous layers, stationary hydrostatic equilibrium, mixing length (for
inclusion of convection), and local thermodynamic equilibrium (LTE)
gradually give way to real physics. The PHOENIX [32] code was developed for
solving non-LTE problems in stellar atmospheres, and the grids of
one-dimensional non-LTE models of Hauschildt et al. \cite{33} and
three-dimensional models (the CO5BOLD code  of Ludwig et al. \cite{46}) were
created for stars of late classes. Three-dimensional hydrodynamic (3DHD)
modeling incorporates significant physical processes, such as the
compressibility of the medium, partial ionization, and nongrey radiation
transfer, but the effective temperature is not an input parameter. This
temperature must be adjusted manually by varying the states of the incoming
gas at the base of the modeled region until the average temperature reaches
the effective temperature value. 3DHD modeling of convection in surface
layers of red giants was carried out using the Optim3D code of Ramirez et
al. \cite{57}. The new BIFROST code intended for 3DHD modeling of the
stellar convection with due account for scattering in the radiative transfer
equation was developed by Hayek et al. \cite{34}. The interaction of
convective motions with stellar magnetic fields is studied by means of the
magnetohydrodynamic (MHD) 3D modeling of the upper part of the convective
region and the photosphere (the MURaM code of  Beeck et al. \cite{14}).
Recently, interesting results of modeling of the hydromagnetic dynamo
processes for the purpose of studying the connection between the generation
of the magnetic flux and its transfer and the distribution of magnetic
fields on the surfaces of cool stars depending on their basic parameters and
rotation were obtained by Isik et al. \cite{37}. In general, the analysis of
3DHD models shows a high level of realism achieved in modeling of the
thermal structure of the photosphere. Unfortunately, the results of 3DHD and
3DMHD modeling of the atmosphere of Arcturus are not yet published.
  \begin{figure}
    \centering
   \includegraphics[width=9.cm]{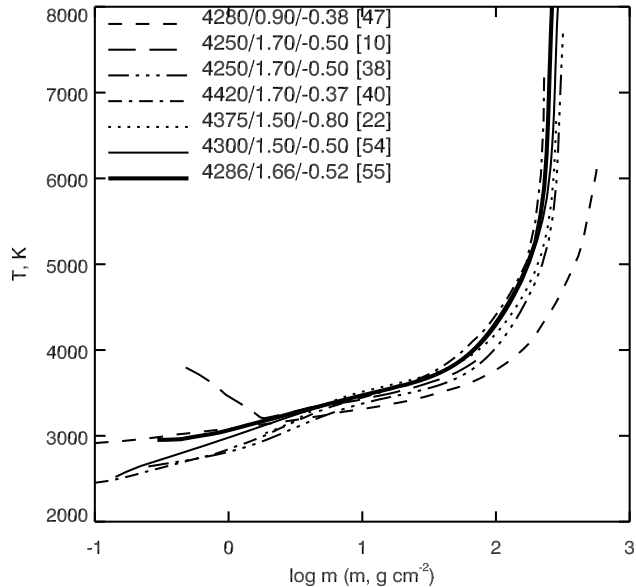}
   \caption
{Dependence  of temperature with depth in the atmosphere of Arcturus
according to  models with different  parameters  $T_{\rm eff}$/$\log
g$/[A/H]: 4280/0.90/$-$0.38 (Mackle and Holweger [47]); 4250/1.70/$-$0.50
(Ayres and  Linsky [10]); 4250/1.70/$-$0.50 (Johnson et al. [38]);
4420/1.70/$-$0.37 (Kipper et al. [40]); 4375/1.50/$-$0.80 (Frisk et al.
[22]); 4300/1.50/$-$0.50 (Peterson et al. [54]); 4286/1.66/$-$0.33 (Ramirez
and  Allende Prieto \cite{56}). }
  \label {Fig2}
  \end{figure}

Classical semiempirical modeling, which is frequently used in studying the
solar atmosphere, may successfully be applied to bright stars if
high-quality spectral observations are available. Semiempirical modeling of
the atmosphere of Arcturus was carried out a number of times. A number of
atmospheric models that present the distribution of temperature and pressure
with height were obtained by Ayres and  Linsky \cite{10}, Frisk et al.
\cite{22}, Johnson et al. \cite{38}, Kipper et al. \cite{40}, Mackle and
Holweger \cite{47}, Peterson et al.~\cite{54} on the basis of the atlas of
the  Arcturus' spectrum (Griffin \cite{25}). Figure 2 shows the temperature
trend of these models as a function of the material mass  $m$ in a column
with unit section lying above the given level ($dm=-\rho dh$, where $\rho$
is the density of the stellar material and $h$ is the geometric height). It
can be seen that the temperature dependences derived in various studies
differ greatly. The temperature difference may reach approximately 500 K at
the level of the continuum formation. The main cause is  the spread of the
values of $T_{\rm eff}$, $\log g$, and [A/H]. The maximum difference in
effective temperatures predicted by the models presented in Fig.~2 amounts
to 170 K, while the maximum differences in the values of surface gravity
acceleration and metallicity amount to 0.8 dex and 0.37 dex, respectively.
Temperature in the deep layers of the photosphere is most sensitive to the
value of $\log g$, while that in the upper layers is most sensitive to the
chemical composition. The temperature trend in the lower chromosphere is
poorly defined due to certain difficulties in modeling these layers (Ayres
and  Linsky \cite{10}). Hence, it appears that the temperature distribution
in the atmosphere of Arcturus is studied insufficiently. New approaches are
needed to define the temperature stratification of the atmosphere of
Arcturus more accurately.

We aim to study the temperature stratification in the photosphere of
Arcturus and construct a semiempirical model based on high-resolution
observations of the broad wings of the H and K Ca II lines.


\section{Advantages of using the wings of the H and K lines for constructing
atmospheric models}
     \label{Advantages }

The idea of using the resonance H ($\lambda$ 396.8 nm) and K ($\lambda$
393.3 nm) lines for constructing atmospheric models was proposed by Linsky
and  Avrett \cite{45} in 1970. They developed a new method of modelling of
solar and stellar phptospheres that we will refer to as ``H-K-modeling'' and
used it for the first time to construct the quiet solar chromosphere. This
method was then applied in studying the active regions of the solar
photosphere (Shine and  Linsky \cite{64}) and the chromospheres of Procyon
(Ayres et al. \cite{11}) and Arcturus (Ayres et al. \cite{10}). The interest
in this method was renewed when new observations and more accurate atomic
data became available. Recently, a new H-K-modeling of the atmosphere in a
sunspot penumbra (Rouppe van der Voort \cite{58}) and the magnetic flux
tubes of the solar faculae (Sheminova et al. \cite{63}) was performed. The
use of the H and K lines in inversion codes (Beck et al. \cite{13}) provides
certain advantages. For example, the need to compute many iron lines is
eliminated, since the broad wings of the H and K lines are formed in a
sufficiently wide height range and cover practically all the layers of the
photosphere. Due to high temperature sensitivity of the Planck function and
the presence of opacity minimum in the continuum around the wavelength
$\lambda = 400.0$~nm, the continuum is formed as deep as it is formed around
$\lambda = 1600.0$~nm. It is of great importance that the wings of the H and
K lines stretching beyond 0.1 nm from their cores are insensitive to non-LTE
effects (Owocki and  Auer \cite{51}). Therefore, they are rather easily
computed under the LTE condition. Due to the great extent of the wings, such
effects as the broadening by micro- and macroturbulence, rotation, and
magnetic fields do not exert marked influence on their profile. These wings
also contain small windows that are free from blends and may be used for
atmospheric modeling. The idea of using multiple blends in the wings of the
H and K lines to determine the radial velocities at different depths in the
photosphere was proposed by Sheminova et al. \cite{63}. The validity and
reliability of H-K-modeling of the solar atmosphere were studied by
Henriques and Kiselman \cite{35} and Sheminova \cite{61}.

\section{Observed profiles of the H and K lines}
     \label{Profiles }

The first spectrophotometric atlas compiled by Griffin \cite{25} was long
used as a data source for spectral studies of the atmosphere of Arcturus due
to its high spectral resolution (150000), broad wavelength range
($\lambda\lambda$~360.0--882.5~nm), low noise level, and insignificant
scattered light. As compared with this first atlas, the second atlas, which
was compiled by Hinkle et al. \cite{36}, offered a wider wavelength range
(120--5300~nm). This range is divided into three parts: the ultraviolet one
(120--380~nm), the optical one (360--930~nm), and the infrared one
(900--5300~nm). In the optical part, the spectral resolution amounts to
approximately 150000, and the signal to noise ratio exceeds 1000. A
comparison of the spectra of the H and K lines taken from these atlases (see
Fig.~3) shows a good agreement between them with the exception of the
regions with maximum absorption in the cores of very strong lines. In the
atlas compiled by Griffin, they are not as deep as in the second atlas. This
difference may be caused by errors in the photometric calibration and the
correction for scattered light. The root mean square (rms) deviation between
the atlas of Griffin \cite{25} (diamonds) and the atlas of Hinkle et al.
\cite{36} (crosses) in the points of the profile that were chosen to be used
in the 1DLTE atmospheric H-K-modeling amounts to 1.3\%.

Unfortunately, an atlas of the absolute radiation flux with a high spectral
resolution is not yet available for Arcturus. Low-resolution measurements of
the absolute flux of Arcturus are nonuniform and vary in their presentation.
For example, the accuracy of absolute measurements in a broad wavelength
range amounts to approximately 3.5\% according to date of Burnashev
\cite{1}, Glushneva et al.  \cite{3}, and Kharitonov et al. \cite{4}. The
available data cannot be used directly for comparison with the synthetic
spectrum. They usually cover certain bands of the spectrum with a typical
width ranging from 2.5 to 5.0~nm. In order to make a comparison with the
results of the computation of the synthetic spectrum, one needs to reduce
the experimental data to monochromatic absolute fluxes and then to absolute
fluxes at the surface of Arcturus taking into account the angular diameter
of the star. Ten series of measurements of the absolute flux of Arcturus
were analyzed by Griffin and  Lynas-Grayin \cite{26}. They concluded that
the systematic errors of the absolute flux measurements may arise due to
neglect of the interstellar absorption, uncertainties of the absolute
calibration, and variations of the UV band radiation flux due to radial
pulsations. In the optical band, different data series show a satisfactory
match, though some measurements differ markedly from the others. In the
infrared band, almost all data series agree well with each other.
  \begin{figure}
    \centering
   \includegraphics[width=14.cm]{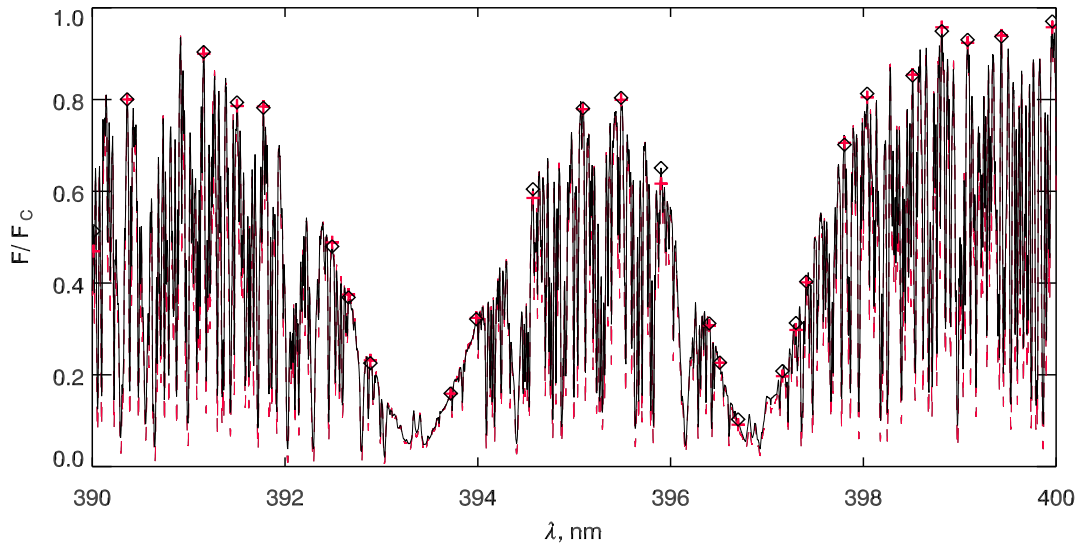}
   \caption
{H and K lines in the spectrum of Arcturus according to the atlases of
Griffin \cite{25} (solid line, diamonds) and  Hinkle et al. \cite{36} (red
dashed line, red crosses). The symbols mark the points of the profile that
were chosen to be used in 1DLTE H-K-modeling of the atmosphere. }
  \label {Fig3}
  \end{figure}

  \begin{figure}
    \centering
   \includegraphics[width=14.cm]{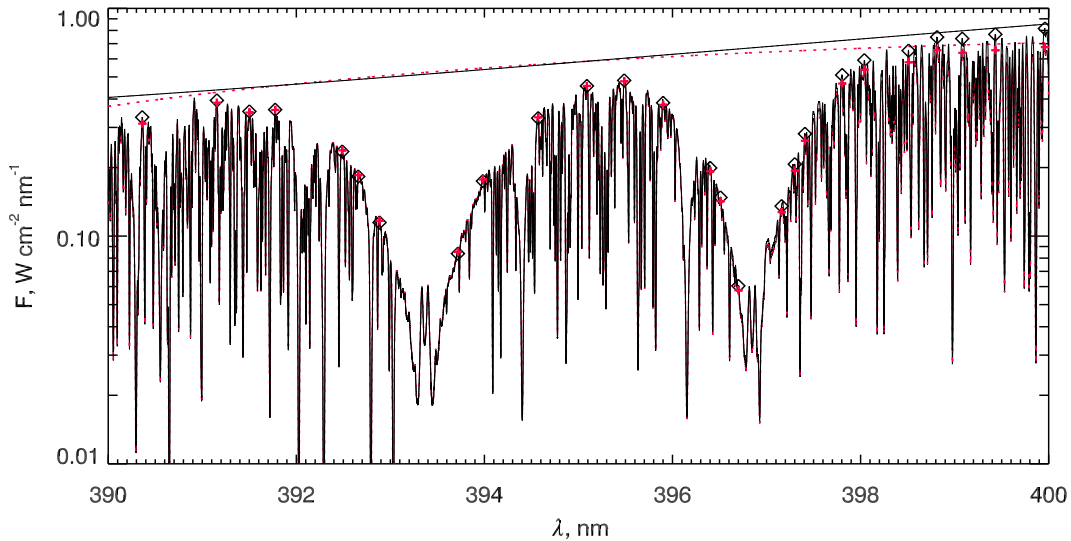}
   \caption
{Absolute flux in the H and K spectrum of Arcturus according to the absolute
calibration of Ayres and Johnson \cite{9} that is based on photometric data
of Fay et al. \cite{20} (solid line, diamonds) and according to the absolute
calibration of Ayres and Linsky \cite{10} based on photometric data of
Willstrop \cite{77} (red dashed line and crosses). The rms deviation between
the indicated calibrations in the chosen points of the profile amounts to
7.2\%. }
  \label {Fig4}
  \end{figure}

If an atlas of the star's spectrum in absolute units is lacking, the problem
of the local continuum arises. In the studies of the stars of late spectral
classes with a dominant picket fence of lines, the local continuum in the
observed spectra is traced in a rough way. Since the procedure of absolute
calibration is nontrivial and time-consuming, the spectral lines measured
relative to the local continuum are given preference when analyzing stellar
atmospheres. Unfortunately, the local continuum cannot be accurately traced
in the spectra of the H and K lines. This spectral region exhibits a
multitude of overlapping atomic and molecular lines.

The results of absolute calibration of  Ayres and Johnson \cite{9} and Ayres
and  Linsky  \cite{10} based on the photometric data of Fay et al. \cite{20}
and Willstrop \cite{77} were used to reduce the spectra of the H and K lines
taken from the atlas  Hinkle et al. \cite{36} to an absolute unit scale.
Figure 4 shows the obtained profiles of radiation flux in the H and K lines
in absolute units. According to  Ayres and  Linsky \cite{10}, the possible
uncertainty of absolute calibration may amount to 20--30\%. This is caused
by the uncertainty of measurements of the angular diameter of Arcturus made
back then and by the possible errors of photometry and determination of the
gauge factors averaged over the photometric bands. Bands with a width of 5
and 3~nm were used by Fay et al. \cite{20} and  Willstrop  \cite{77},
respectively. Figure 4 shows that the maximum difference (5--10\%) between
the calibration results based on two different sets of photometric data is
developed in the far wing of the H line. Since this difference is smaller
than the uncertainty of the given calibration, it is considered
insignificant according to Ayres and Johnson \cite{9}, Ayres and  Linsky
\cite{10}.

\section{Calculation of the wings of the H and K Ca II lines}
     \label{Calculation}

The SPANSAT  software (Gadun and  Sheminova \cite{2}) was used for
calculating the synthetic spectrum. This software was modified so as to
allow calculation of the spectrum containing any number of overlapping
spectral lines. The atomic parameters of all (approximately 6000) lines in
the wavelength range of 390--400 nm were taken from the VALD (Kupka et al.
\cite{42}) database. Molecular lines were not included in the list of blends
that was used in calculating the wings. According to Johnson et al.
\cite{38}, Arcturus is sufficiently hot for the molecules not to play a
significant role in the structure of its atmosphere. The only exceptions may
be the CO molecules, which produce a cooling effect near the surface, and
the TiO molecules, which produce a slight heating of the entire atmosphere.
Building on these results, one might assume that the contribution of
molecular lines to the intensity of the wings of the H and K lines is
insignificant. In order to check this, we carried out test calculations
using the full list of molecular lines of Tsymbal \cite{72}. The results of
this test are presented in Fig.~5. In the outermost wings, the incorporation
of molecular lines slightly reduces the relative  flux. This is especially
evident in the shortwave wing of the K line. The 26 points (chosen to be
used in H-K-modeling and marked by diamonds and crosses in Fig.~5) in the
wings of the H and K lines are virtually insusceptible to the influence of
molecular blends.
  \begin{figure}
    \centering
   \includegraphics[width=14.cm]{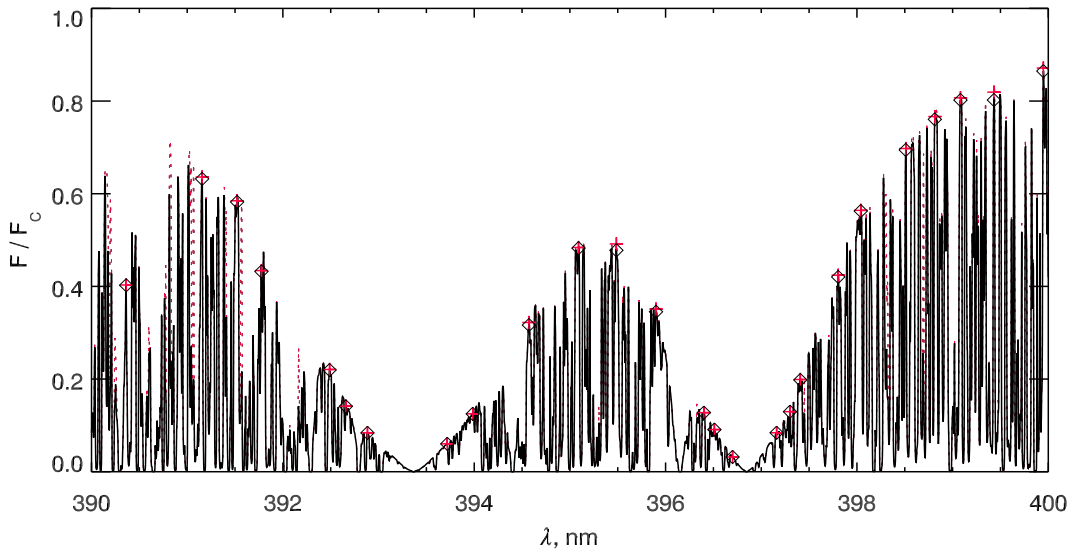}
   \caption
{H and K spectrum of Arcturus calculated with due account for molecular
lines (solid line, diamonds) and without regard to them (red dashed line and
crosses). The correction for the UV deficit was not introduced. The rms
deviation in the chosen points of the profile amounts to 0.5\%. }
  \label {Fig5}
  \end{figure}

The Arcturus photospheric velocities field, whose parameters must be defined
in order to calculate the wings of the H and K lines, is best described by
an anisotropic radial-tangential model of micro- and macroturbulent
velocities according to  Gray \cite{23} and Sheminova and  Gadun \cite{62}.
It is also known that the amplitude of the microturbulent velocity increases
with height from 0.5 to 2.2 km/s in the region of $\log \tau_5$ varying from
0 to $-2$ (Takeda \cite{69}). Let us remark here that microturbulence
decreases with height in the specified optical depth range on the Sun, and
the transition to increasing with height occurs only around $\log \tau_5
=-3$ (Gurtovenko and  Sheminova \cite{28}). In the atmospheres of stars of
late classes, the velocity field is closely associated with convective
motions. It appears that the opposite behaviors of turbulent motions in the
atmospheres of Arcturus and the Sun are conditioned by differing properties
of convective motions that penetrate the photosphere. The results of
modeling of the convective envelope of Arcturus (Kuker and Rudiger
\cite{41}) in the mean field approximation show that the outer convective
zone of Arcturus extends depthward to a distance from the star's center that
equals 3\% of its radius. Convective motions resembling solar granulation
and supergranulation take place throughout the whole interior of the star.
Since the influence that the velocity field exerts on the broad wings of the
H and K lines is insignificant, the amplitude of the total velocities field
$V_t$ is used in the present work to simplify the calculations. This
amplitude was introduced by Ayres and Johnson  \cite{9} in the form of
$V_t^2=V_{\rm mic}^2+V_{\rm mac}^2$. In the present work, a value of
$V_t=3.7$~km/s is adopted on the assumption that isotropic $V_{\rm mic}$ and
$V_{\rm mac}$ are (according to Sheminova and Gadun \cite{62}) equal to 1.6
and 3.3~km/s, respectively.

According to  Brown et al. \cite{15} and Gray and  Brown \cite{24}, the most
recent and accurate value ($V\sin i=1.5 \pm0.3$~km/s) of the rotation
velocity of the atmosphere of Arcturus is smaller than the previous one,
$V\sin i=2.4$~km/s (Gray \cite{23}). The equatorial rotation velocity of
Arcturus equals $1.8 \pm 0.3$~km/s, the angle of inclination of its axis to
the line of sight equals $58 \pm 25^\circ$, and roman the rotation period
amounts to approximately two years. We adopt  $V\sin i=1.5 \pm0.3$~km/s for
calculating the profile of the wings.

For the Sun as a star, the amplitude of the total turbulent velocities field
$V_t$ is taken to be equal to 2.6~km/s ($V_{\rm mic} = 1.2$~km/s and $V_{\rm
mac} = 2.3$~km/s according to Sheminova \cite{5}), and the rotation velocity
$V \sin i$ is assumed to take on a standard value of 1.8~km/s.

The damping parameters play an important role in calculations of the
synthetic spectrum. The shape of the broad wings of the H and K lines is
determined primarily by radiation damping and pressure effects that arise
due to collisions between absorbing calcium ions and neutral hydrogen and
helium atoms (van der Waals broadening). Therefore, the $C_6$ van der Waals
broadening constant becomes an important parameter in calculations of the
wings of the H and K lines. Ayres \cite{8} showed that a correction factor
must be applied to the value of $C_6$, which is calculated according to the
classic Unsold formula, when studying the H and K lines. This correction
factor is equal to $1.6 \pm 0.5$ in the case of the H and K lines in the
center of the solar disk.

Figure 6 shows the profiles of the H and K lines calculated for Arcturus and
the Sun (as a star) using different values of the correction factor to
$C_6$. At present, it is no longer necessary to determine the correction
factor for each spectral line. A technique of Barklem and  O'Mara \cite{12}
based on quantum-mechanical estimates  may be used to determine a
sufficiently reliable value of $C_6$. We use a code that was obtained
through the courtesy of the authors of this technique to calculate $C_6$.

In the calculations of spectral lines, the elemental abundances correspond
to the values that were used in the initial models.

  \begin{figure}
    \centering
   \includegraphics[width=13.cm]{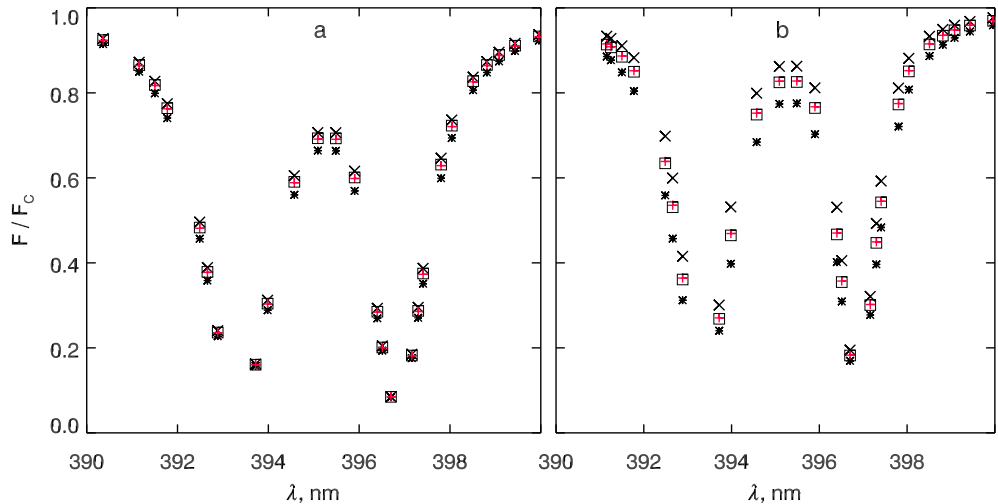}
   \caption
{Synthetic H and K lines (without blends). The calculations were carried out
for Arcturus (left panel) and the Sun as a star (right panel) using
different values of the damping constant: 1.0$C_6$ (crosses), 1.3$C_6$
(squares, left panel), 1.4$C_6$ (squares, right panel), and 2.0$C_6$
(asterisks). Red pluses denote the van der Waals broadening determined using
the technique presented by Barklem and O'Mara \cite{12}. }
  \label {Fig6}
  \end{figure}

\section{Deficit in UV absorption}
     \label{Deficit}

The calculated radiation flux of Arcturus in the UV band is significantly
higher than the observed one. This discrepancy became known as the deficit
in UV absorption (Gustafsson and \cite{29}), since it is natural to assume that
the effect is caused by some unknown absorbers that may significantly reduce
the UV flux similarly to the continuous absorption. This UV deficit may be
produced by a multitude of unknown weak lines (line haze) or by unknown
sources of continuous absorption. The influence of non-LTE effects,
atmospheric inhomogeneities, invisible companions of the star, and errors of
the absolute measurements of the UV spectrum due to atmospheric absorption
is also possible.

Ignoring the UV absorption deficit in modeling of the stellar atmosphere
leads to construction of a cooler model. Therefore, this effect is
considered important. According to our estimates (Sheminova et al.
\cite{63}), in the case of the solar H and K spectrum in the center of the
disk, the discrepancy between the results of calculations and observations
in the UV band is rather small and depends on the photospheric model. This
discrepancy was estimated using a correction factor $f(\lambda)$ to the
opacity $ \kappa_{\lambda}$ in the continuum. This factor was included in
the calculation of profiles in the form of $ \kappa_{\lambda}^{\rm new}=
f(\lambda) \kappa_{\lambda}$. In the well-known HOLMUL, VALC, and HSRA-SP
solar models, the correction factor averaged over the range of
$\lambda\lambda$~390--400~nm equals roughly $f \approx 1.14$, 1.05, and
1.01, respectively. The hotter a model of the solar photosphere, the larger
is the correction factor.

Detailed studies  of the deficit in UV absorption by Short et al. \cite{65,
66, 67, 68} suggest that large differences between the calculated and
observed fluxes in the UV band are most likely produced by the unaccounted
continuous absorption of the diatomic molecules of metal hydrides (MgH, SiH,
and FeH). Generally speaking, the discovery of the deficit in UV absorption
on Arcturus came as no surprise because a similar solar deficit was already
known. Examination of all available results of absolute measurements of the
solar radiation flux revealed that newer and more accurate measurements
reduce the solar deficit in UV absorption, which equaled 10\% according to
Neckel \cite{49} and  Neckel and  Labs  \cite{50}. Measurements that were
carried out at the Peak Terskol Observatory by Burlov-Vasiljev et al.
\cite{16} and measurements of the EURECA satellite (Thuillier et al.
\cite{70}) showed that the solar UV flux is 8\% and 4\% higher compared to
tha date of Neckel \cite{49} and  Neckel and  Labs~\cite{50}.

Short and Hauschildt \cite{66} estimate that the correction factor to UV
absorption in the spectrum of the Sun as a star $f$ equals roughly 1.15,
whereas in the spectrum of Arcturus $f \approx 2$. These estimates were
produced taking into account a comprehensive list of absorption lines, the
sphericity of atmosphere, non-LTE effects, a more accurate chemical
composition, and the inhomogeneity of atmosphere. Introduction of all these
corrections did not reduce the $f$-value for Arcturus. The authors believe
that a significant additional opacity is most probably linked to continuous
absorption by unknown sources. It was also discovered that Arcturus is the
only star with an anomalously large deficit in UV absorption in the studied
group of stars of G--K classes with metallicity ranging from 0 to $-0.5$.

Thus, an analysis of the results of the studies of the deficit in UV
absorption revealed that this effect is big in Arcturus' spectrum  and its
nature  is still unknown. In the future, we plan to check the assumption
that the companion of Arcturus may be responsible for the excess of
radiation flux in the star's UV spectrum. In the present paper, missing
sources of continuous absorption in the UV band are taken into account by
using the correction factor $f(\lambda)$ to the opacity in continuum.

\section{Method of modeling of the stellar atmospheres}
     \label{Method}

The process of semiempirical modeling begins with the selection of an
initial model that determines the initial approximation for the distribution
of temperature and pressure with height. This model may be selected from a
grid of standard stellar models, which are calculated using the given values
of $T_{\rm eff}$, $\log g$, and metallicity. Other parameters of the model
that are needed to calculate the synthetic spectrum are computed on the
assumption of a hydrostatic equilibrium in the vertical direction in the
case of plane-parallel geometry. According to Short and Hauschildt \cite{66},
the spherical model of
the atmosphere of Arcturus is only slightly (by 50 K) warmer in the outer
atmospheric layers than the plane-parallel one. Therefore, in order to
simplify the calculations, the sphericity of atmosphere is not taken into
account in the present work.

When the initial model is prepared, the wings of the H and K lines are
calculated with due regard to all blends. In order to speed up the
calculations and compare their results with observations, the radiation flux
is calculated in 26 chosen points that are spread over the entire region of
the H and K spectrum and are almost free from blends. These points of the
profile are denoted by symbols in Figs.~3--8. The synthetic profile is then
compared to observational data in 26 points. In order to get a quantitative
estimation of the best possible fit between the profiles, the root mean
square deviation of the synthetic spectrum from the observed one is
calculated in those same points according to the following formula:

\begin{equation}\label{eq_1}
  \sigma(\frac {F_{\rm calc} - F_{\rm obs}}{F_{\rm obs}}) =
  \sqrt{\frac{1}{n}\sum_{i=1}^n \frac{(F_{\rm calc}(\lambda_i) -
  F_{\rm obs }(\lambda_i))^2}
  {F_{\rm obs }(\lambda_i)^2}},
\end{equation}
where $ F_{\rm calc} (\lambda_i) $ and $ F_{\rm obs} (\lambda_i) $ are the
calculated and the observed absolute radiation fluxes in line for a
wavelength of $\lambda_i$ ($i=1, ... , 26$). If fitting of the profiles is
carried out on a relative scale, $\sigma $ is equal to
\begin{equation}\label{eq_2  }
  \sigma(r_{\rm calc} - r_{\rm obs}) =\sqrt{\frac{1}{n}\sum_{i=1}^n
  (r_{\rm calc} (\lambda_i)-r_{\rm obs} (\lambda_i))^2},
\end{equation}
where $r (\lambda_i) $ is the ratio of the radiation flux in line to the
radiation flux in local continuum for a certain point $\lambda_i$ of the
profile.

If the calculated and the observed spectra do not show a satisfactory
agreement, the temperature distribution is modified until the agreement
becomes optimal. Each successive change of the temperature distribution is
performed manually in 13 specially selected points at different heights
starting from the upper layers. If the effective temperature of the initial
model is to be kept unchanged, the correction of temperature is concluded in
the continuum formation layers ($\log\tau_5\approx1$). This makes it
possible to refine the temperature trend keeping the value of $T_{\rm eff}$
fixed. The points where the temperature is interactively changed are
selected so that their height distribution is uneven. Their selection
depends on the magnitude of the temperature gradient. An interpolation over
all height points specified in the initial model is then carried out, and
gas pressure and other model parameters are recalculated in compliance with
the new temperature stratification.

Each successive iteration in the process of matching the calculated spectrum
with the observed one reduces the rms deviations $\sigma$. In principle,
such a procedure is burdened by the problem of uniqueness of solution. Our
past experience in constructing models for the solar photosphere
(Sheminova~\cite{61}, Sheminova et al.~\cite{63}) tells us that a fitting
procedure described above produces satisfactory results. It is usually
assumed that the solar atmosphere model that offers the best possible match
between the theory and observations as a result of the fitting represents
true solar atmosphere within the limits of assumptions inherent to the
method and averaging intrinsic to observations. In the case of other stars,
the situation becomes slightly more complicated. Stellar profiles with their
low spatial resolution do not offer an adequate representation of a quiet
atmosphere because of large-scale stellar surface inhomogeneities. Since
spectral lines include contributions from the entire surface, a
multicomponent model is generally required. However, it may be expected that
a single-component model would suffice in the case of Arcturus and
solar-type stars. It was shown earlier by Sheeley \cite{60} that the solar
Ca II integrated profiles (and their wings in particular) undergo only
slight changes in relation to cyclic solar activity. The errors of absolute
calibration of observations due to problems associated with photometry of
weak objects are also common to stellar spectra. Therefore, even the best
possible match does not make it possible to construct a model of a real
stellar atmosphere.

The SPANSAT software is used to recalculate the model at each iteration. The
coefficient of continuous absorption is recalculated on a temperature
stratification change according to the ATLAS [43] algorithm taking into
account the following sources of absorption and scattering: (1) H~I; (2)
H$_2^{+}$; (3) bound-free and free-free transitions of H$^û$; (4) Rayleigh
scattering by H~I; (5) He~I; (6) bound-free and free-free transitions of
He~II; (7) free-free transitions of He$^-$; (8) Rayleigh scattering by He~I;
(9) bound-free and free-free transitions of C~I, Mg~I, Si~I, and Al~I
(low-temperature absorbers) and Si~II, Mg~II, Ca~II, N~I, and O~I
(medium-temperature absorbers); (10) bound-free transitions of C~II--IV,
N~II--V, O~II--VI, and Ne~I--II (high-temperature absorbers); and (11) Rayleigh
scattering by electrons and H$_2$. It should be noted that all model
computations and calculations of spectra are carried out with the chemical
composition that is specified in the initial model.

In order to make sure that the used method of H-K-modeling of stellar
atmospheres is reliable, we applied it first to the Sun as a star and then
to Arcturus.

\section{Initial conditions of modeling}
     \label{Initial}

The HSRA-SP-M (Sheminova et al.~\cite{63}) model served as an initial model
in the case of the Sun. Chemical composition of the solar atmosphere
corresponds to  data of Fontenla \cite{21}. An observed spectrum of the Sun
as a star presented in electronic form in absolute units was kindly
furnished by R. Rutten. This spectrum agrees with data from the atlas
compiled by Neckel and Labs~\cite{50} with an absolute calibration carried
out by Neckel~\cite{49}.

For Arcturus the values of effective temperature ($T_{\rm eff} = 4286$~K),
surface gravity acceleration ($\log g = 1.66$), and metals' (C, O, Na, Mg,
Al, Si, S, K, Ca, Ti, Cr, Mn, Fe, and Ni) abundance relative to hydrogen
were taken from paper of Ramirez and  Allende Prieto \cite{56}.
It should be
noted that we obtained the average value of [A/H]  equals $\rm[A/H]=
\log(\frac{N_{A}}{N_{H}})_{\ast}-\log(\frac{N_{A}}{N_{H}})_{\odot}=-0.33 $,
whereas the PASTEL database lists a value of [A/H]~$ = -0.56$. The abundance
of alpha elements (O, Mg, Si, S, Ca, and Ti) $[\alpha/\rm Fe]=
\lg(\frac{N_{\alpha}}{N_{Fe}})_{\ast}
-\lg(\frac{N_{\alpha}}{N_{Fe}})_{\odot}=0.34 $ \cite{56}, while the
abundance of iron [Fe/H] equals $-0.52$.  The
temperature and pressure stratification depending on mass $m$ was calculated
based on these data using the modified ATLAS12 code of  Pavlenko \cite{52}.
The initial model of Arcturus is
designated as 4286/1.66/$-$0.33 in accordance with the adopted fundamental
parameters ($T_{\rm eff}$/$\log g$/[A/H]) values.

Three additional initial models from the Kurucz grid were used to examine
the influence of uncertainties in the values of the fundamental parameters
of Arcturus on the end result. The metal abundance value in these models was
obtained simply by subtracting 0.5 dex from the solar abundance used in the
Kurucz grid (Anders and Grevesse \cite{7}), i.e., [A/H]~$ = -0.5$. The
effective temperature $T_{\rm eff}$ was equal to 4250 and 4300~K, whereas
the surface gravity acceleration $\log g$ was equal to 1.5 and 2.0. These
models are designated as $4250/1.5/{-}0.5$, $4250/2.0/{-}0.5$, and
$4300/1.5/{-}0.5$.

  \begin{figure}
    \centering
   \includegraphics[scale=1.]{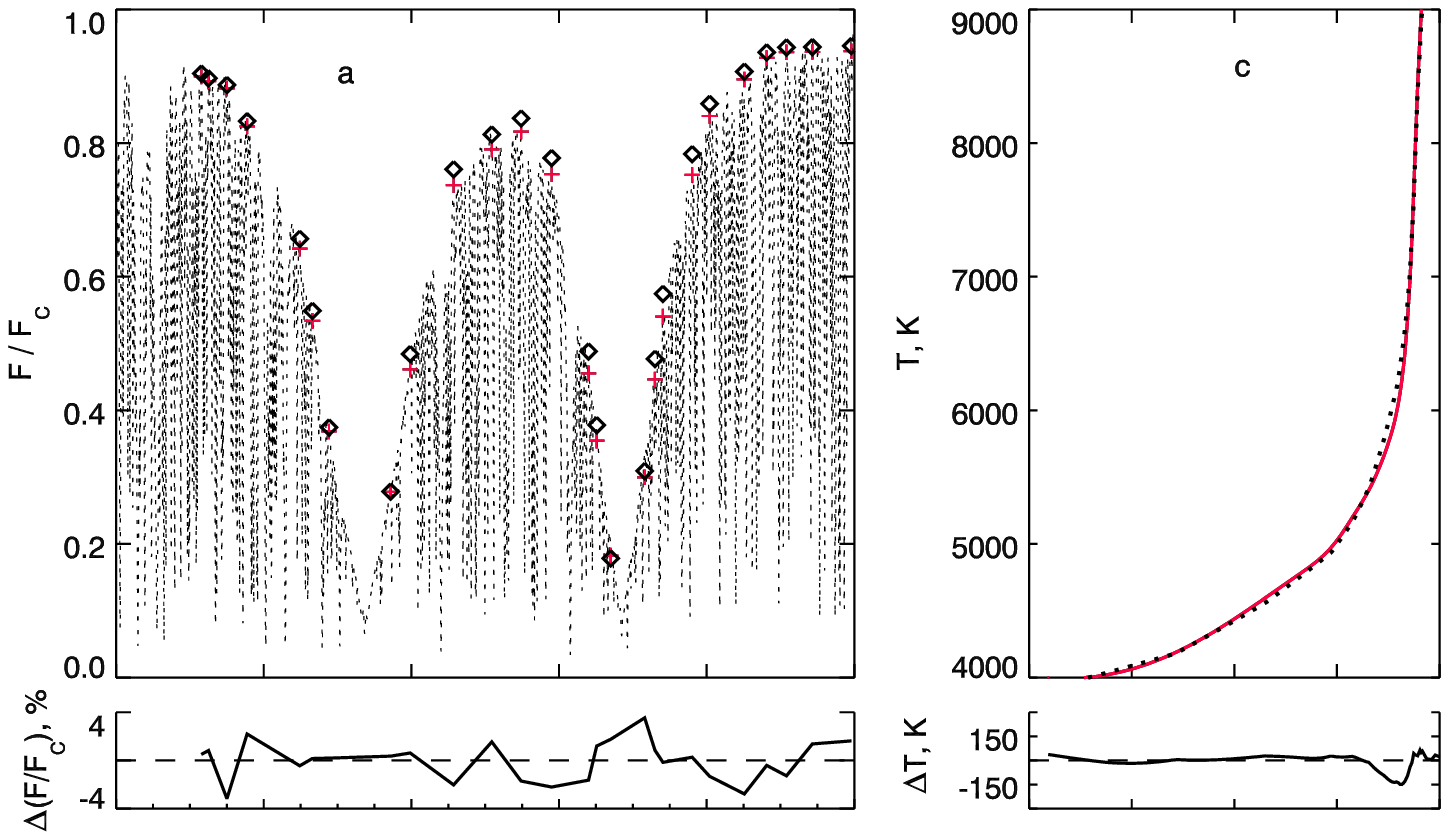}
    \includegraphics[scale=1.]{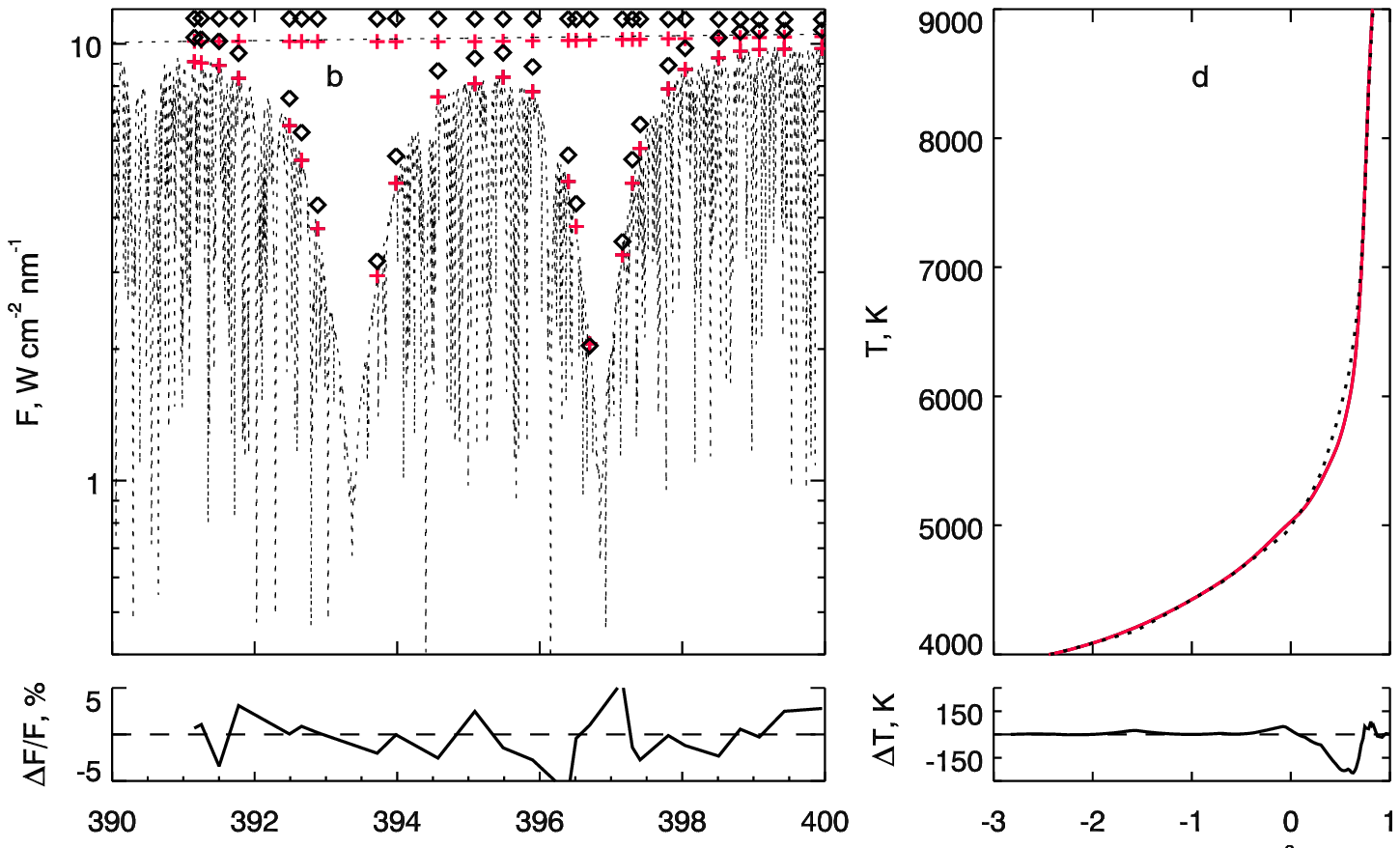}
   \caption
{Results of 1DLTE modeling of the Sun as a star. The observed radiation flux
(dashed line) is presented in (a) relative and (b) absolute units. Diamonds
denote the profile calculated using the initial model without correction for
the UV deficit. The best match between the calculated (red crosses) and the
observed profiles is shown in the chosen 26 points, and the differences
between them are shown on the lower panels. The obtained temperature
stratification (right panels) is denoted by a red  solid line in comparison
with the initial HSRA-SP-M model (dashed line). The differences between them
are shown on the lower panels.}
  \label {Fig7}
  \end{figure}

\section{Results}
     \label{Results}

\subsection{Modeling of the solar atmosphere (the Sun as a star)}

Though a high-precision spectral atlas in absolute units made by
Neckel \cite{49} and  Neckel and  Labs  \cite{50} is
available for the Sun as a star, we also carried out H-K-modeling of the
solar atmosphere using a spectrum in relative units in order to compare the
modeling results.

Figure 7a shows that the profiles of the H and K lines calculated using the
initial HSRA-SP-M model without corrections to the continuum (diamonds) only
slightly deviate from the observed ones. The best match between the
calculation results (crosses) and observations was achieved by introducing
the correction factor to the coefficient of continuous absorption and
slightly (by 150 K) reducing temperature in the deep photospheric layers
(see Fig.~7c). The correction factor was needed only for the shortwave wing
of the K line. Matching of the wings of the profiles produced the values of
the minimal rms deviation $\sigma= 0.016$ and the correction factor
$f(\lambda)= 1.15$, 1.10, 1.00, 1.00, and 1.00 corresponding to $\lambda =
390.0$, 392.5, 395.0, 398.0, and 400.0 nm, respectively.

Figures 7b and 7d present the results of atmospheric modeling with the use
of the H and K spectrum in absolute units. The synthetic flux in the H and K
lines  and in the continuum calculated  without including the correction
factor to the continuum with the initial model clearly exceeds the observed
one (see Fig.~7b). The average deficit in UV absorption for the calculated
continuum amounts to 9.3\% in the wavelength range of 390--400 nm. The best
match between the synthetic and observed spectra was achieved at a minimal
root mean square deviation $\sigma= 0.015$, the correction factor
$f(\lambda)= 1.20$, 1.20, 1.20, 1.17, and 1.15 corresponding to $\lambda =
390.0$, 392.5, 395.0, 398.0, and 400.0 nm, respectively, and at reduction of
temperature by 150--250 K in the layers of effective continuum formation.
The obtained photospheric model of the Sun as a star turned out to be cooler
than the HSRA-SP-M model inferred in the same way for the center of the
solar disk (see Fig.~7d).

It should be noted that the local continuum level in the observed spectrum
of the H and K lines agrees satisfactorily with the calculated one (see
Fig.~7b). For this reason, the synthetic profiles of the wings of the H and
K lines in relative units ($F_L/F_C$) calculated without including the
correction to the continuum turned out to be close to the observed profiles.
If the observed continuum matched perfectly with the calculated one, there
would be no need to introduce the correction factor to the continuum on a
relative scale since the UV deficit would simply be compensated.

\subsection{Modeling of the atmosphere of Arcturus}

  \begin{figure}
    \centering
   \includegraphics[scale=1.]{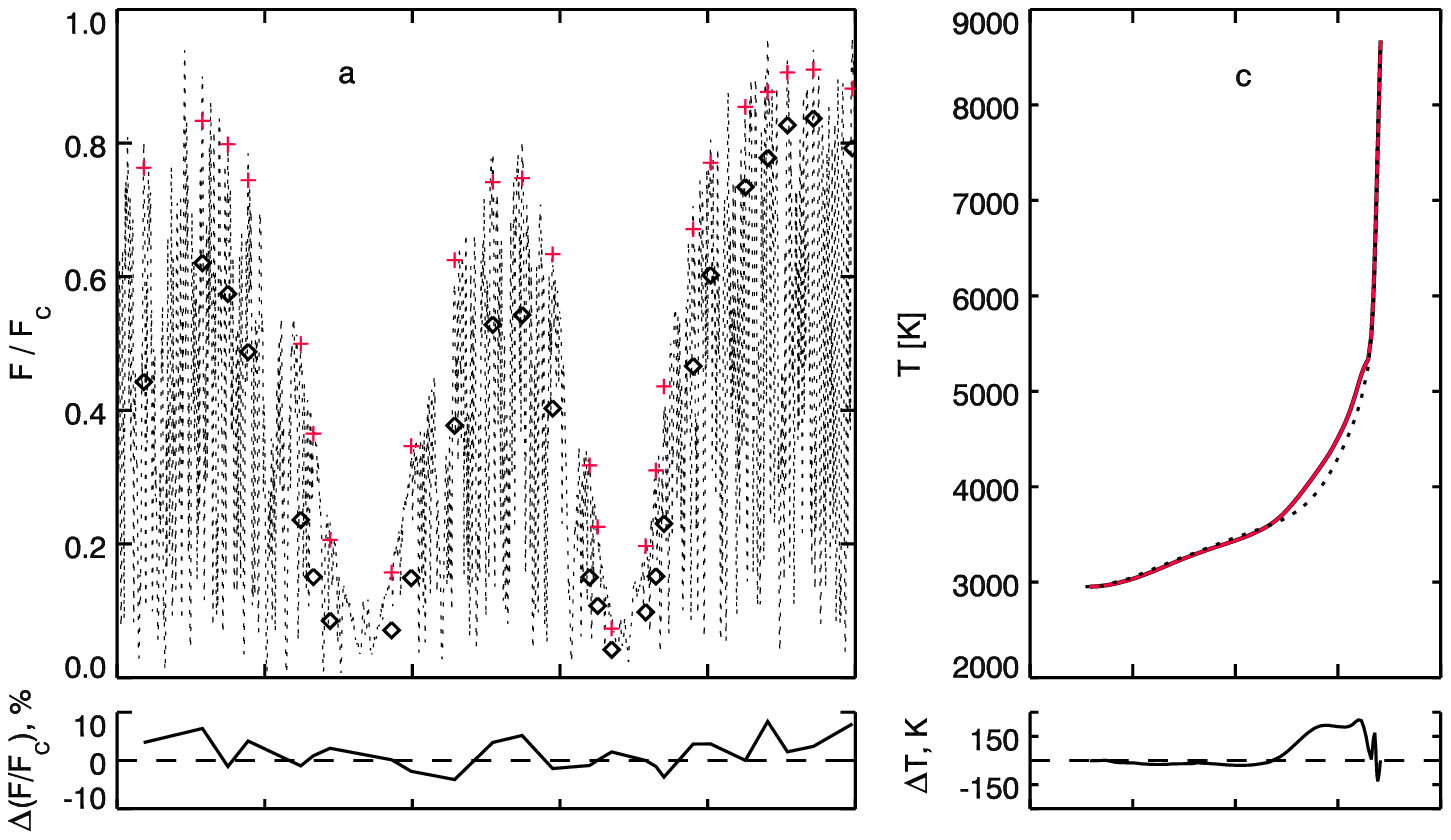}
   \includegraphics[scale=1.]{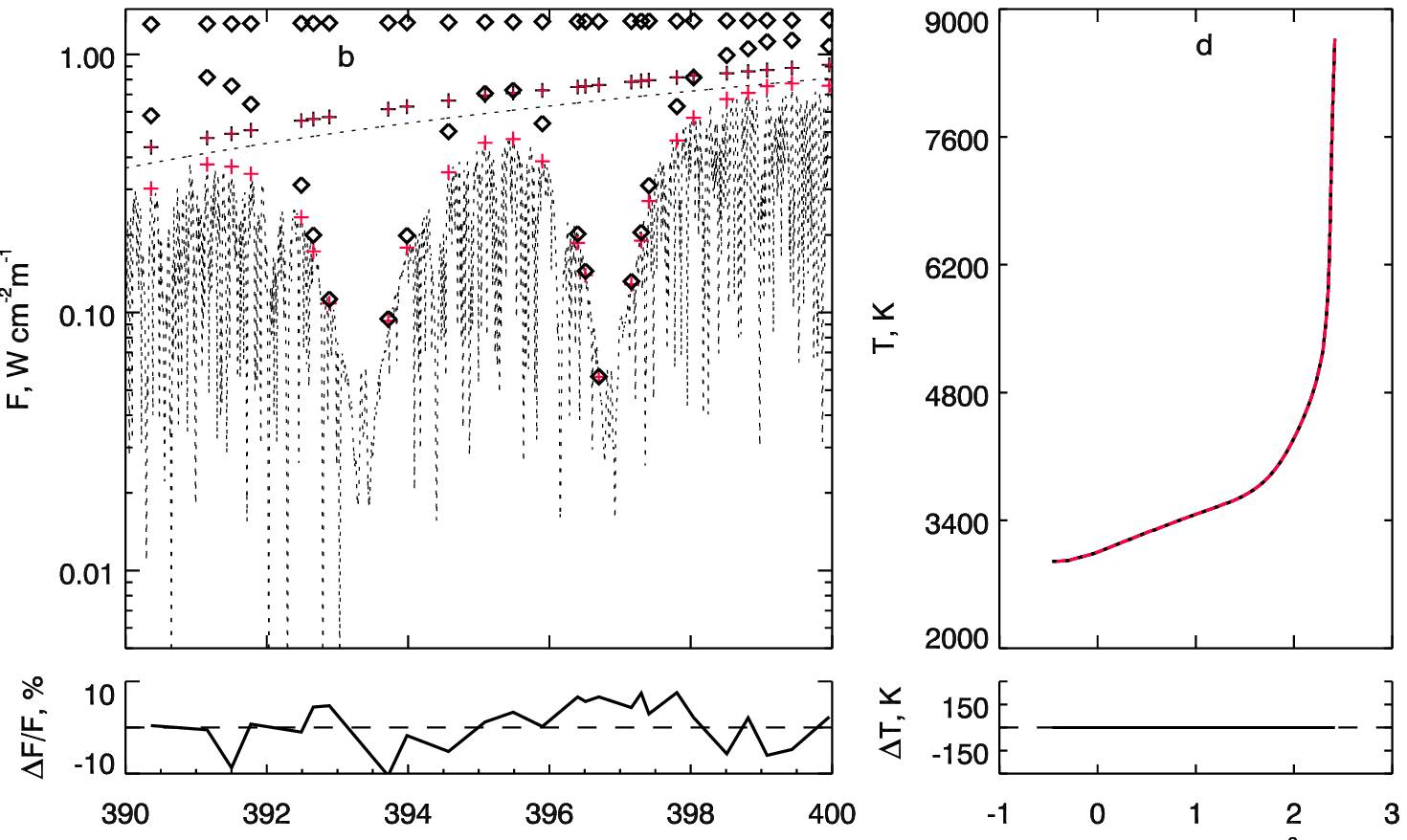}
   \caption
{Results of 1DLTE modeling of Arcturus with the $4286/1.66/{-}0.33$ initial
model (dashed line). For key see Fig.~7.}
  \label {Fig8}
  \end{figure}

Figure 8a shows the results of matching of the profiles in relative units
using the initial model atmosphere  with 4286/1.66/$-$0.33. The synthetic
profiles (diamonds) calculated without including the correction to the
continuum ($f = 1$) are deeper than the observed ones not only in the wings
but also in the central parts of lines. Inclusion of the correction factor
$f(\lambda)$ in the calculations of opacity and variation of the temperature
trend with depth provided a significantly closer match between the synthetic
(crosses) and the observed profiles. However, satisfactory results were
still not achieved. The correction to the opacity turned out to be rather
large compared to the solar data. In the case of Arcturus, $f(\lambda)$ was
found to be equal to 3.16, 2.62, 2.26, 1.99, and 1.81 corresponding to
$\lambda  = 390.0$, 392.5, 395.0, 398.0, and 400.0 nm, respectively. The
minimal rms deviation $\sigma$ was twice as large as in the case of the
solar spectrum and equaled 0.035. Figure 8c shows that the obtained model
turned out to be hotter by 200 K in deep photospheric layers than the
4286/1.66/$-0.33$ initial model.

Other initial models of armosphere  such as 4250/1.5/$-0.5$,
4250/2.0/$-0.5$, 4300/1.5/$-0.5$ as well as 4260/0.9/$-0.38$ (Mackle and
Holweger~\cite{47}), 4250/1.7/$-0.50$ (Johnson et al.~\cite{38}),
4420/1.7/$-0.37$ (Kipper et al. \cite{40}), 4375/1.5/$-0.80$ (Frisk et
al.~\cite{22}), and 4300/1.5/$-0.5$ (Peterson~\cite{54}) did not produce
significantly better matches between the calculated and the observed
profiles in relative units. The value of $\sigma  = 0.035$ remained the
lowest. These results showed that the differences in the initial conditions
of modeling do not exert virtually any influence on the results of matching
of the calculated profile with observations. The most likely cause of
matching of the profiles being unsatisfactory is the combined influence of
uncertainties of the local continuum in atlas of  Hinkle et al.~\cite{36}
and of the UV deficit.

  \begin{figure}
    \centering
   \includegraphics[width=14cm]{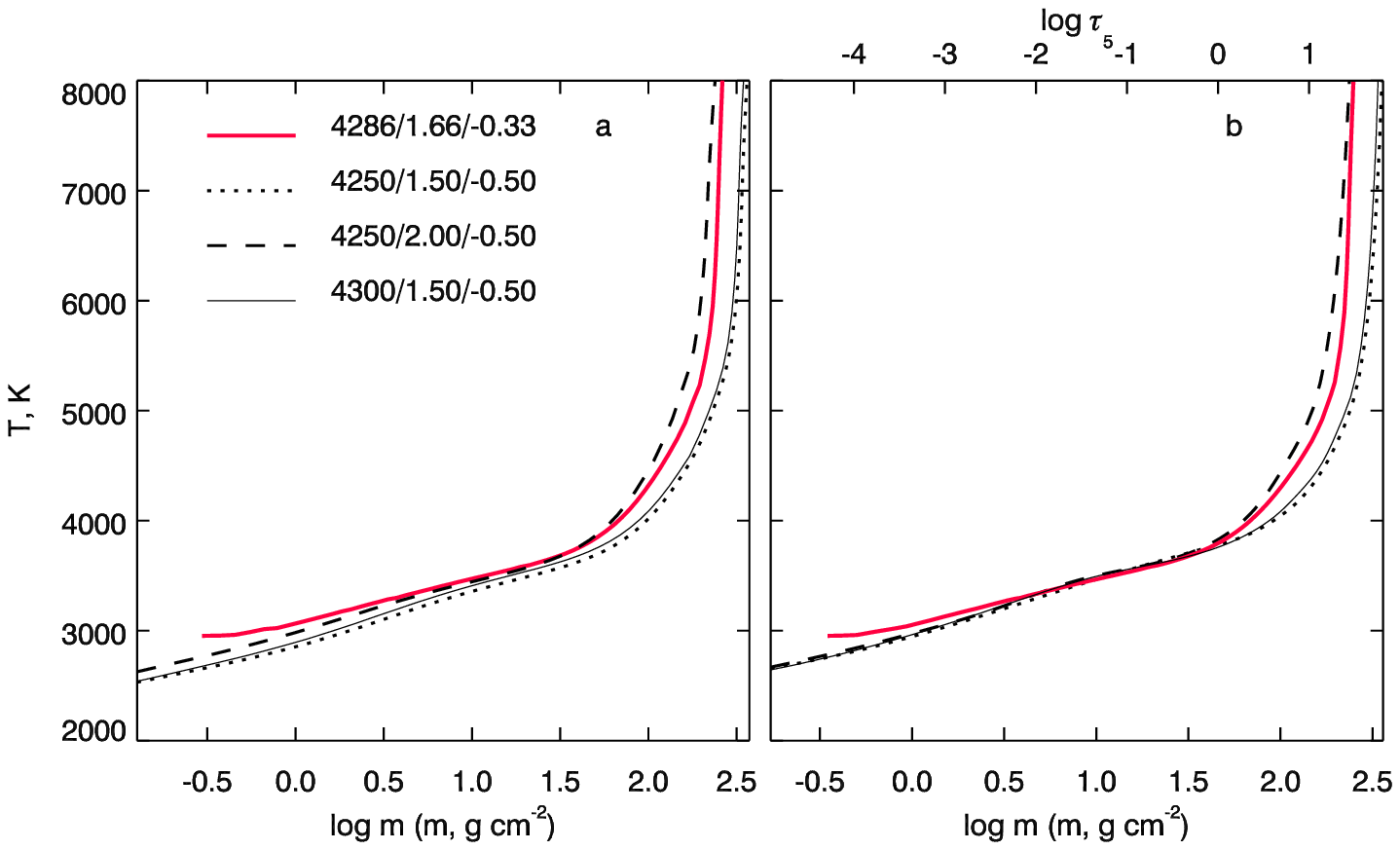}
   \caption
{Distribution of temperature with logarithm of mass $\log m$ in the
atmosphere of Arcturus as described by (a) the initial models and (b) the
models obtained by fitting the absolute radiation fluxes in the wings of the
H and K lines. A scale of optical depths $\log \tau_5$ is shown above
right
 .}
  \label {Fig9}
  \end{figure}

The synthetic flux in absolute units (diamonds) calculated using the
4286/1.66/$-$0.33 initial model (without the correction to the continuum)
significantly exceeds the observed flux in the H and K lines (see Fig.~8b).
This is indicative of a very large deficit in UV absorption on Arcturus.
According to our estimates, this deficit is vastly superior to the solar one
and amounts to 42.6\%. Inclusion of unknown absorption in the continuum
using the factor $f(\lambda)$ made it possible to produce a satisfactory
match between the synthetic spectrum (crosses) and observations with
$\sigma= 0.028$. In the process, the values of $f(\lambda) = 2.45$, 1.95,
1.65, 1.45, and 1.35 (photometry of Fay et al.~\cite{20}) and $f(\lambda) =
2.20$, 1.90, 1.70, 1.55, and 1.45 (photometry of   Willstrop \cite{77})
corresponding to $\lambda = 390.0$, 392.5, 395.0, 398.0, and 400.0 nm,
respectively, were obtained. We attempted to produce a better match between
the profiles (i.e., obtain a $\sigma$ value that would be equal to the solar
one) by varying the temperature trend, but failed. The likely causes of
failure are an insufficiently reliable absolute calibration, the fact that
the resolution of the observed spectrum of Arcturus is lower than the
resolution of the solar spectrum, and probable presence of a companion of
Arcturus.

The obtained results of matching the spectra in absolute units show that the
model atmosphere 4286/1.66/$-$0.33  gives a sufficiently reliable
description of the observed wings of the H and K lines without correcting
the temperature trend (but including the correction for missing absorption
in the continuum) and indicates underestimating of the local continuum by an
average of 12\%.

Figure 9a and 9b show temperature stratifications of a few initial models
and the stratifications obtained through H-K-modeling. It is evident that,
in the region of optical depths $\log \tau_5$ ranging from $-2.5$ to 0.3
where the wings of the H and K lines are effectively formed, new temperature
stratifications come close to the 4286/1.66/$-$0.33 model. The correction
factors to the continuum vary from case to case. In the case of the
4250/1.5/$-$0.5 model, $f (\lambda) = 1.96$, 1.60, 1.42, 1.33, and 1.30; in
the case of the 4250/2.0/$-$0.5 model, $f (\lambda) = 2.05$, 1.76, 1.59,
1.47, and 1.38; and in the case of the 4300/1.5/$-$0.5 model, $f (\lambda) =
2.1$, 1.8, 1.6, 1.5, and 1.5. The higher the value of $T_{\rm eff}$, the
greater is the UV deficit.

\section{Discussion}

The results of matching of the synthetic and the observed solar radiation
flux in the wings of the H and K lines are presented in Figs. 7a and 7b in
relative and absolute units and are almost identically good ($\sigma  =
0.016$ and 0.015) because the local continuum is sufficiently well-traced.
This is evidenced by a satisfactory ($\sigma = 0.014$) agreement between the
observed and the calculated (with account for the correction for the UV
deficit; see Fig.~7b) continua within the limits of the errors of current
analysis. It is for this very reason that the new distributions of
temperature with height inferred for the Sun as a star are almost identical
(see Figs. 7c and 7d).

  \begin{figure}
    \centering
   \includegraphics[width=15.cm]{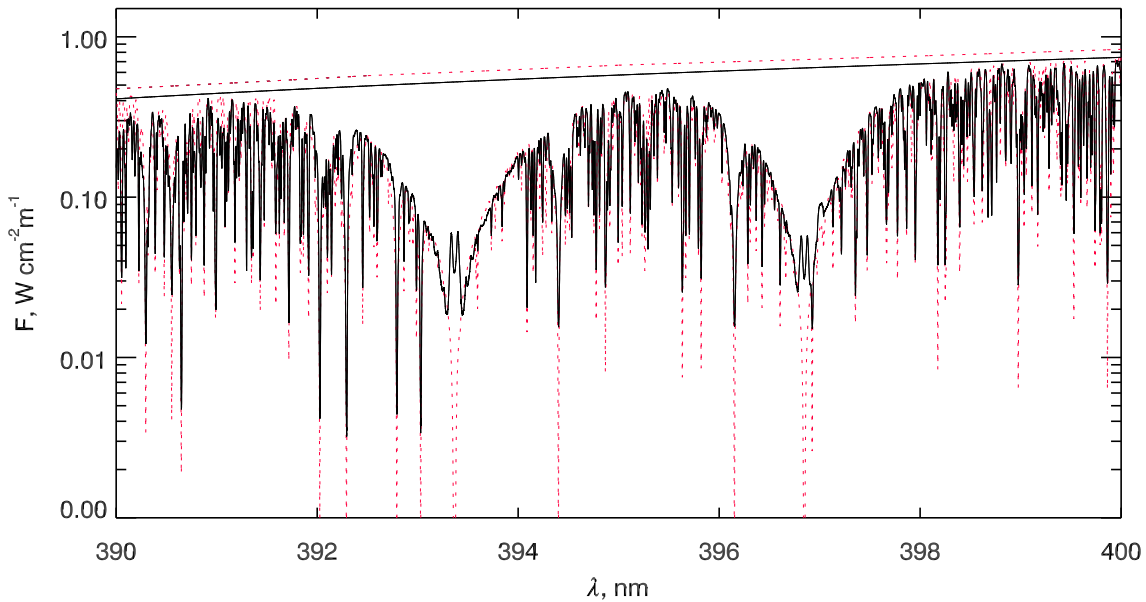}
   \caption
{Synthetic (red dashed line) and observed (Hinkle et al. \cite{36}, solid
line) H-K-spectra of Arcturus with all (approximately 6000) blends on an
absolute unit scale. The 4286/1.66/$-$0.33 model is used;  $V_{\rm mic}$ =
1.6 km/s, $V_{\rm mac}$ = 3.3 km/s, and $V \sin i = 1.5$~km/s.}
  \label {Fig10}
  \end{figure}

The model atmosphere derived for the Sun as a star in the present work
turned out to be cooler by 100--200 K in the layers of  continuum formation
than the HSRA-SP-M  model (Sheminova et al.~\cite{63}) based on the the H
and K line observations at the solar disk centre. The most likely causes of
this difference are the horizontal and vertical temperature inhomogeneities
that develop due to penetrating convection in the surface atmospheric layers
and are not taken into consideration in one-dimensional modeling. In this
respect, it is interesting to note that the analysis of influence of
temperature inhomogeneities on the results of one-dimensional modeling
(Sheminova~\cite{61},  Uitenbroek and  Criscuoli~\cite{74}) showed that the
models of the solar photosphere  derived from the observations in the center
of the disk may produce an overestimation of 200 K in the deep layers. In
the case of one-dimensional modeling  on the base of integrated solar
spectrum, the temperature inhomogeneities associated with granulation
produce differing effects on the emergent intensity in the central region
and in the peripheral regions of the solar disk. The flux from the disk
centre produces an overestimation of temperature, while the peripheral flux
produces an underestimation of temperature. These effects compensate each
other  and the  constructed model atmosphere  turns out to be closer to
reality than a one-dimensional model atmosphere obtained from observations
at the solar disk centre.

In the case of Arcturus, the results of matching of the synthetic and the
observed radiation flux in the wings of the H and K lines in relative
(Fig.~8a) and absolute (Fig.~8b) units are indicative of major issues in
both the calculation of the continuum and the tracing of the local continuum
in the observed H-K-spectrum. According to our results, the local continuum
is underestimated by an average of 12\%. It is for this very reason that a
satisfactory agreement with observations could not be reached on a relative
units scale ($\sigma = 0.035$). The reduction of continuum by 12\% leads to
a 200~K increase in temperature in the layers of continuum formation (see
Fig.~8c) when using relative radiation fluxes for atmospheric modeling.
Since the results of matching the synthetic H-K-spectrum with the observed
one on an absolute scale are not dependent on the local continuum, they are
better ($\sigma = 0.028$) than the results on a relative scale but worse
than the results for the solar spectrum ($\sigma = 0.015$). This accentuates
the importance of using radiation fluxes in absolute units when modeling
stellar atmospheres. In the contrary case, the errors due to inaccurate
local continuum arise inevitably.

It is worth noting that the 4286/1.66/$-0.33$ initial model (see Fig.~8d)
produces the closest (without additional variations of temperature
stratification) description of the observed H-K-spectrum in absolute units
(see Fig.~8b). This means that the values of the fundamental parameters of
Arcturus obtained  by Ramirez and Allende Prieto \cite{56} are reliable.
Supposing the 4286/1.66/$-0.33$ model is closest to reality, the results of
matching obtained using the set of other initial models (see Fig.~9a)
demonstrate the influence of possible errors in the fundamental parameters
on the atmospheric temperature trend. The obtained novel distributions of
temperature with height (see Fig.~9b) agree with each other in the layers of
effective formation of the H and K wings (from $\log \tau_5\approx -2.5$ to
$\log \tau_5\approx 0.3$). This points to the fact that the method used
produces a reliable estimate of temperature stratification in the
photospheric layers regardless of the uncertainties of the fundamental
parameters. In deeper subphotospheric layers, where the used method is
insensitive
%
 \begin{table}

  \begin{center}{

{\small

{Model of the atmosphere of Arcturus (4286/1.66/$-$0.33) }\\
 \vspace{0.25cm}
 \begin{tabular}{cccccc}
 \hline
  $ H$  cm& $m$, g cm$^{-2}$& $T$, K & $P_g$, Pa & $P_e$, Pa & $\rho$, g cm$^{-3}$\\
 \hline

 2.95345E+10 &3.48036E-01 & 2952.6 & 1.57168E+01 &1.53279E-04 & 8.19207E-11 \\
 2.79759E+10 &5.02573E-01 & 2961.0 & 2.27783E+01 &2.01092E-04 & 1.18388E-10 \\
 2.59219E+10 &8.12300E-01 & 3020.5 & 3.69195E+01 &3.45085E-04 & 1.88105E-10 \\
 2.46580E+10 &1.08678E+00 & 3067.6 & 4.94521E+01 &4.96510E-04 & 2.48091E-10 \\
 2.25721E+10 &1.73921E+00 & 3155.9 & 7.92471E+01 &9.22029E-04 & 3.86441E-10 \\
 2.11051E+10 &2.40289E+00 & 3215.3 & 1.09564E+02 &1.38996E-03 & 5.24410E-10 \\
 1.96139E+10 &3.31782E+00 & 3276.5 & 1.51363E+02 &2.08954E-03 & 7.10940E-10 \\
 1.81157E+10 &4.56336E+00 & 3329.9 & 2.08276E+02 &3.03616E-03 & 9.62566E-10 \\
 1.66278E+10 &6.23055E+00 & 3385.5 & 2.84460E+02 &4.39340E-03 & 1.29306E-09 \\
 1.58901E+10 &7.25724E+00 & 3412.5 & 3.31378E+02 &5.25038E-03 & 1.49442E-09 \\
 1.51564E+10 &8.43632E+00 & 3438.4 & 3.85262E+02 &6.23910E-03 & 1.72433E-09 \\
 1.44265E+10 &9.78865E+00 & 3463.4 & 4.47065E+02 &7.37831E-03 & 1.98650E-09 \\
 1.37005E+10 &1.13370E+01 & 3487.5 & 5.17830E+02 &8.68522E-03 & 2.28504E-09 \\
 1.29778E+10 &1.31086E+01 & 3511.3 & 5.98795E+02 &1.01949E-02 & 2.62440E-09 \\
 1.22586E+10 &1.51317E+01 & 3535.0 & 6.91259E+02 &1.19402E-02 & 3.00934E-09 \\
 1.15430E+10 &1.74379E+01 & 3559.2 & 7.96663E+02 &1.39741E-02 & 3.44462E-09 \\
 1.08311E+10 &2.00604E+01 & 3584.8 & 9.16523E+02 &1.63771E-02 & 3.93457E-09 \\
 1.01212E+10 &2.30491E+01 & 3607.2 & 1.05312E+03 &1.90119E-02 & 4.49289E-09 \\
 9.41463E+09 &2.64387E+01 & 3635.6 & 1.20804E+03 &2.23451E-02 & 5.11355E-09 \\
 8.71230E+09 &3.02689E+01 & 3666.4 & 1.38310E+03 &2.63407E-02 & 5.80536E-09 \\
 8.01484E+09 &3.45803E+01 & 3701.1 & 1.58015E+03 &3.12284E-02 & 6.57037E-09 \\
 7.32191E+09 &3.94196E+01 & 3741.0 & 1.80133E+03 &3.73034E-02 & 7.41015E-09 \\
 6.63237E+09 &4.48419E+01 & 3786.4 & 2.04915E+03 &4.48659E-02 & 8.32850E-09 \\
 5.94497E+09 &5.09061E+01 & 3838.8 & 2.32631E+03 &5.43984E-02 & 9.32586E-09 \\
 5.25775E+09 &5.76817E+01 & 3898.9 & 2.63598E+03 &6.64218E-02 & 1.04043E-08 \\
 4.91296E+09 &6.13668E+01 & 3933.1 & 2.80440E+03 &7.36897E-02 & 1.09730E-08 \\
 4.56746E+09 &6.52602E+01 & 3968.8 & 2.98234E+03 &8.17448E-02 & 1.15642E-08 \\
 4.21934E+09 &6.93923E+01 & 4008.1 & 3.17119E+03 &9.09569E-02 & 1.21759E-08 \\
 3.86874E+09 &7.37726E+01 & 4049.8 & 3.37138E+03 &1.01267E-01 & 1.28112E-08 \\
 3.51550E+09 &7.84148E+01 & 4094.0 & 3.58354E+03 &1.12785E-01 & 1.34703E-08 \\
 3.15586E+09 &8.33826E+01 & 4142.2 & 3.81058E+03 &1.25900E-01 & 1.41570E-08 \\
 2.79303E+09 &8.86492E+01 & 4192.3 & 4.05127E+03 &1.40360E-01 & 1.48715E-08 \\
 2.42144E+09 &9.43146E+01 & 4246.6 & 4.31019E+03 &1.56753E-01 & 1.56195E-08 \\
 2.04269E+09 &1.00379E+02 & 4303.5 & 4.58735E+03 &1.74932E-01 & 1.64041E-08 \\
 1.65711E+09 &1.06863E+02 & 4363.0 & 4.88368E+03 &1.95037E-01 & 1.72255E-08 \\
 1.25706E+09 &1.13929E+02 & 4427.6 & 5.20658E+03 &2.17941E-01 & 1.80964E-08 \\
 8.51372E+08 &1.21456E+02 & 4494.1 & 5.55056E+03 &2.43160E-01 & 1.90064E-08 \\
 4.29373E+08 &1.29681E+02 & 4566.1 & 5.92648E+03 &2.72339E-01 & 1.99735E-08 \\
 0.00000E+00 &1.38473E+02 & 4642.5 & 6.32825E+03 &3.06174E-01 & 2.09769E-08 \\
 -4.38065E+08 &1.47894E+02 & 4724.1 & 6.75879E+03 &3.47136E-01 & 2.20167E-08 \\
 -8.77946E+08 &1.57802E+02 & 4817.0 & 7.21158E+03 &4.01559E-01 & 2.30386E-08 \\
 -1.31291E+09 &1.68050E+02 & 4917.4 & 7.67988E+03 &4.77011E-01 & 2.40337E-08 \\
 -1.72702E+09 &1.78182E+02 & 5036.4 & 8.14287E+03 &5.97737E-01 & 2.48802E-08\\
 -2.11111E+09 &1.87890E+02 & 5144.9 & 8.58645E+03 &7.57385E-01 & 2.56819E-08 \\
 -2.46709E+09 &1.97201E+02 & 5257.9 & 9.01190E+03 &9.88801E-01 & 2.63746E-08 \\
 -3.01119E+09 &2.11645E+02 & 5573.6 & 9.67161E+03 &2.14671E+00 & 2.66995E-08\\
 -3.40244E+09 &2.22087E+02 & 5892.0 & 1.01483E+04 &4.69394E+00 & 2.64952E-08 \\
 -3.66785E+09 &2.29050E+02 & 6275.5 & 1.04657E+04 &1.12515E+01 & 2.56382E-08 \\
 -3.84486E+09 &2.33523E+02 & 6663.9 & 1.06691E+04 &2.50490E+01 & 2.45822E-08\\
 -3.97953E+09 &2.36793E+02 & 6995.2 & 1.08175E+04 &4.65522E+01 & 2.36968E-08 \\
 -4.11573E+09 &2.39995E+02 & 7305.6 & 1.09627E+04 &7.93749E+01 & 2.29266E-08 \\
 -4.17852E+09 &2.41436E+02 & 7428.8 & 1.10281E+04 &9.69853E+01 & 2.26454E-08\\
 -4.25064E+09 &2.43072E+02 & 7564.7 & 1.11023E+04 &1.20268E+02 & 2.24034E-08 \\
 -4.35261E+09 &2.45354E+02 & 7743.4 &  1.12058E+04 &1.57571E+02 & 2.20183E-08 \\
 -4.45569E+09 &2.47625E+02 & 7904.7 & 1.13088E+04 &1.99071E+02 & 2.16891E-08\\
 -4.72696E+09 &2.53447E+02 & 8279.6 & 1.15729E+04 &3.30933E+02 & 2.09536E-08 \\
 -4.90552E+09 &2.57180E+02 & 8487.0 & 1.17423E+04 &4.29900E+02 & 2.05695E-08 \\
 -5.07393E+09 &2.60640E+02 & 8673.4 & 1.18993E+04 &5.37551E+02 & 2.02151E-08\\
\hline
\end{tabular}
} }
\end{center}

\end{table}
to temperature variations, the temperature trend fits with the initial
models. Here, the discrepancies between the initial models are determined
mainly by differences in the value of surface gravity acceleration. Since
electrons are supplied mostly by hydrogen in the deep layers, the influence
of the chemical composition there is insignificant. In the uppermost layers
of the photosphere and the lower chromosphere, where $\log \tau_5 <-2.5$,
surface gravity acceleration and chemical composition of the atmosphere
exert equal influence on the temperature trend. The main sources of free
electrons in these layers are metals. The 4286/1.66/$-$0.33 initial model
adopts a metal abundance [A/H] value that exceeds the values adopted by
other models by an average of 0.17 dex. The increase in metal abundance
leads to an increase in electron pressure and, therefore, to an increase in
opacity, reduction of gas pressure, and a temperature increase in these
layers. Therefore, the temperature of the 4286/1.66/$-$0.33 initial model in
the upper layers is higher than the temperature of other initial models (see
Fig.~9a). The difference between the obtained models in the upper layers is
evidently lessened (see Fig.~9b) as a result of H-K-modeling, but it is not
eliminated. In order to refine the model in these chromospheric layers, one
must calculate the profiles of the cores of the H and K lines, but this lies
beyond the scope of the current analysis.

Figure 10 presents the synthetic H-K-spectrum calculated with uniform step
$\Delta\lambda =5$~m\AA\, in the framework of the 4286/1.66/$-$0.33 model,
including the correction to opacity in the continuum. The cores of all
strong lines turned out to be deeper than the observed ones due to the used
LTE approximation. The H$\varepsilon$ (397~nm) Balmer line in the wing of
the H line does not exhibit the observed emission. According to Ayres and
Linsky \cite{10} the emission is caused by the fact that the conditions of
generation of this line are highly sensitive to a temperature rise in the
chromosphere. The calculations of this line and the cores of the H and K
lines make it possible to refine the chromospheric temperature
stratification, but this requires solving a non-LTE problem and taking into
account the sphericity of the atmosphere of Arcturus.

\section{Conclusions}

The one-dimensional modeling  of the atmosphere of Arcturus was performed
based on the synthesis  of the extended wings of the H and K Ca II lines and
the minimization of differences between the calculations and the
observations. An important feature of this  modeling  is the requirement of
using the star's radiation flux presented in absolute units. If this
requirement is met, no significant restrictions to applying the used method
to stellar spectra are found. If the radiation flux relative to the
continuum is used for modeling, one needs to obtain a reliable estimate of
the local continuum. Errors associated with visual tracing of the local
continuum usually lead to its underestimation, and this results in a
temperature excess in the layers of effective formation of the continuum.
Our results show that the usage of the stellar spectrum in relative units
leads to a 200 K (this value exceeds the spread of estimated values of
$T_{\rm eff}$ for Arcturus) increase in temperature in the layers of
continuum formation. This is preconditioned by underestimation of the local
continuum by an average of 12\% in the atlases of Arcturus and by the
presence of unaccounted opacity sources or processes that lead to
redistribution of radiation across the spectrum.

The UV deficit is still a topical problem of constructing the synthetic
spectrum of Arcturus in the region of the H and K Ca II lines. We tried
compensating for the missing opacity by using a correction factor and found
that the value of correction to the continuous spectrum depends on
wavelength and effective temperature of the model. This value decreases
significantly with an increase in wavelength and goes up when $T_{\rm eff}$
increases. The synthetic continuum (without corrections) exceeds the
observed one by an average of 42.6\% in the case of Arcturus, and this
difference equals 9.3\% in the case of the Sun. It is possible to produce a
reliable estimate of the correction to the continuum by using absolute
fluxes and matching only the far wings of the H and K lines. In the case of
Arcturus, we recommend using the correction factor to the coefficient of
continuous absorption $f (\lambda) = 2.20$, 1.90, 1.70, 1.55, and 1.45 for
the wavelengths of  $\lambda = 390.0$, 392.5, 395.0, 398.0, and 400.0 nm,
respectively. The average correction factor  equals 1.8. It agrees
satisfactorily with the data of Short and Hauschildt \cite{66}.

The semiempirical model atmosphere inferred by 1DLTE H-K-modeling agrees
with the theoretical model calculated  for the fundamental parameters of
Arcturus $T_{\rm eff} = 4286$~K, $\log g = 1.66$, [Fe/H] = $-$0.52, and
[A/H]~=~$-0.33$ (Ramirez and Allende Prieto \cite{56}). This allows us to
conclude that the fundamental parameters of Arcturus derived by Ramirez and
Allende Prieto \cite{56} are reliable. We recommend using the
4286/1.66/$-$0.33 model for spectral analysis of Arcturus in further
studies.  The model parameters (geometrical height $H$, material mass $m$,
temperature $T$, gas pressure $P_g$, electron pressure $P_e$, density
$\rho$)  are presented in Table.

The obtained results of 1DLTE H-K-modeling of the atmosphere of Arcturus
testify to the fact that, at present, it is not possible to derive a
temperature stratification of the stellar atmosphere that would be more
accurate than the one inferred by standard method that was used in the
creation of the grids of model atmospheres. The main reason for this is the
lack of   high-resolution stellar spectra calibrated to absolute fluxes with
high quality. In the future, it would be worthwhile to conduct a similar
study of the atmosphere of Arcturus using calibrated spectral observations
carried out with a resolution of 500000. In principle, such spectra may
already be obtained using the 2-meter telescope of the Peak Terskol
Observatory.

{\bf Acknowledgements.} I am grateful to A. Shavrina for calculation of the
synthetic spectrum with account for molecular lines, to Ya. Pavlenko for
calculation of the theoretical model  of the atmosphere of
Arcturus with the $T_{\rm eff}=4286$~K, $\log g=1.66$ and {A/H]~$=-0.33$.
I am so thankful  to the reviewer for helpful comments and suggestions that
improved the presentation of results of the study.



\end{document}